\documentclass[aps,prl,reprint,amsmath,amssymb,superscriptaddress]{revtex4-2}
\usepackage{graphicx}
\usepackage{dcolumn}
\usepackage{bm}
\usepackage[colorlinks,linkcolor=blue,anchorcolor=blue,citecolor=blue,urlcolor=blue,filecolor=blue,menucolor=blue,runcolor=blue]{hyperref}

\usepackage[english]{babel}

\begin{document}

\title{Observation of hidden altermagnetism in Cs$_{1-\delta}$V$_2$Te$_2$O}

\author{Guowei Yang}
\thanks{These authors contributed equally to this paper.}
\affiliation{School of Physics, Zhejiang University, Hangzhou 310058, China}
\affiliation{Center for Correlated Matter, Zhejiang University, Hangzhou 310058, China}

\author{Ruihan Chen}
\thanks{These authors contributed equally to this paper.}
\affiliation{School of Physics, Zhejiang University, Hangzhou 310058, China}
\affiliation{Center for Correlated Matter, Zhejiang University, Hangzhou 310058, China}

\author{Changchao Liu}
\thanks{These authors contributed equally to this paper.}
\affiliation{School of Physics, Zhejiang University, Hangzhou 310058, China}

\author{Jing Li}
\thanks{These authors contributed equally to this paper.}
\affiliation{School of Physics, Zhejiang University, Hangzhou 310058, China}

\author{Ze Pan}
\affiliation{School of Physics, Zhejiang University, Hangzhou 310058, China}
\affiliation{Center for Correlated Matter, Zhejiang University, Hangzhou 310058, China}

\author{Teng Hua}
\affiliation{School of Physics, Zhejiang University, Hangzhou 310058, China}
\affiliation{Center for Correlated Matter, Zhejiang University, Hangzhou 310058, China}

\author{Pengyue Xiong}
\affiliation{School of Physics, Zhejiang University, Hangzhou 310058, China}
\affiliation{Center for Correlated Matter, Zhejiang University, Hangzhou 310058, China}

\author{Liwei Deng}
\affiliation{National Key Laboratory of Materials for Integrated Circuits, Shanghai Institute of Microsystem and Information Technology, Chinese Academy of Sciences, Shanghai 200050, China}

\author{Naifu~Zheng}
\affiliation{State Key Laboratory of Quantum Functional Materials and Department of Physics, Southern University of Science and Technology, Shenzhen 518055, China}

\author{Yu Tang}
\affiliation{School of Physics, Zhejiang University, Hangzhou 310058, China}
\affiliation{Center for Correlated Matter, Zhejiang University, Hangzhou 310058, China}

\author{Hao Zheng}
\affiliation{School of Physics, Zhejiang University, Hangzhou 310058, China}
\affiliation{Center for Correlated Matter, Zhejiang University, Hangzhou 310058, China}

\author{Weifan Zhu}
\affiliation{School of Physics, Zhejiang University, Hangzhou 310058, China}
\affiliation{Center for Correlated Matter, Zhejiang University, Hangzhou 310058, China}

\author{Yifu Xu}
\affiliation{School of Physics, Zhejiang University, Hangzhou 310058, China}
\affiliation{Center for Correlated Matter, Zhejiang University, Hangzhou 310058, China}

\author{Xinying Zheng}
\affiliation{School of Physics, Zhejiang University, Hangzhou 310058, China}
\affiliation{Center for Correlated Matter, Zhejiang University, Hangzhou 310058, China}

\author{Xin~Ma}
\affiliation{School of Physics, Zhejiang University, Hangzhou 310058, China}
\affiliation{Center for Correlated Matter, Zhejiang University, Hangzhou 310058, China}

\author{Xiaoping Wang}
\affiliation{Neutron Scattering Division, Neutron Sciences Directorate, Oak Ridge National Laboratory, Oak Ridge, Tennessee 37831, USA}

\author{Shengtao Cui}
\affiliation{National Synchrotron Radiation Laboratory, University of Science and Technology of China, Hefei 230029, China}

\author{Zhe Sun}
\affiliation{National Synchrotron Radiation Laboratory, University of Science and Technology of China, Hefei 230029, China}

\author{Zhengtai Liu}
\affiliation{Shanghai Synchrotron Radiation Facility, Shanghai Advanced Research Institute, Chinese Academy of Sciences, Shanghai 201210, China}

\author{Mao Ye}
\affiliation{Shanghai Synchrotron Radiation Facility, Shanghai Advanced Research Institute, Chinese Academy of Sciences, Shanghai 201210, China}

\author{Chao Cao}
\affiliation{School of Physics, Zhejiang University, Hangzhou 310058, China}
\affiliation{Center for Correlated Matter, Zhejiang University, Hangzhou 310058, China}
\affiliation{Institute for Advanced Study in Physics, Zhejiang University, Hangzhou 310058, China}

\author{Ming Shi}
\affiliation{School of Physics, Zhejiang University, Hangzhou 310058, China}
\affiliation{Center for Correlated Matter, Zhejiang University, Hangzhou 310058, China}
\affiliation{Institute for Advanced Study in Physics, Zhejiang University, Hangzhou 310058, China}

\author{Lunhui Hu}
\affiliation{School of Physics, Zhejiang University, Hangzhou 310058, China}
\affiliation{Center for Correlated Matter, Zhejiang University, Hangzhou 310058, China}

\author{Qihang Liu}
\affiliation{State Key Laboratory of Quantum Functional Materials and Department of Physics, Southern University of Science and Technology, Shenzhen 518055, China}

\author{Shan Qiao}
\email{qiaoshan@mail.sim.ac.cn}
\affiliation{National Key Laboratory of Materials for Integrated Circuits, Shanghai Institute of Microsystem and Information Technology, Chinese Academy of Sciences, Shanghai 200050, China}

\author{Guanghan Cao}
\email{ghcao@zju.edu.cn}
\affiliation{School of Physics, Zhejiang University, Hangzhou 310058, China}
\affiliation{Institute for Advanced Study in Physics, Zhejiang University, Hangzhou 310058, China}
\affiliation{Interdisciplinary Center for Quantum Information, and State Key Laboratory of Silicon and Advanced Semiconductor Materials, Zhejiang University, Hangzhou 310058, China}

\author{Yu Song}
\email{yusong$\_$phys@zju.edu.cn}
\affiliation{School of Physics, Zhejiang University, Hangzhou 310058, China}
\affiliation{Center for Correlated Matter, Zhejiang University, Hangzhou 310058, China}

\author{Yang Liu}
\email{yangliuphys@zju.edu.cn}
\affiliation{School of Physics, Zhejiang University, Hangzhou 310058, China}
\affiliation{Center for Correlated Matter, Zhejiang University, Hangzhou 310058, China}
\affiliation{Institute for Advanced Study in Physics, Zhejiang University, Hangzhou 310058, China}

\date{\today}%
\addcontentsline{toc}{chapter}{Abstract}
	
\begin{abstract}
    Altermagnets are characterized by anisotropic band/spin splittings in momentum space, dictated by their spin-space group symmetries. However, the real-space modulations of altermagnetism are often neglected and have not been explored experimentally. Here we combine neutron diffraction, angle-resolved photoemission spectroscopy (ARPES), spin-resolved ARPES and density functional theory to demonstrate that Cs$_{1-\delta}$V$_2$Te$_2$O realizes a spatially modulated form of altermagnetism, i.e., hidden altermagnetism. Such a state in Cs$_{1-\delta}$V$_2$Te$_2$O results from its G-type antiferromagnetism and two-dimensional electronic states, allowing for the development of spatially alternating altermagnetic layers, whose local spin polarizations are directly verified by spin-resolved ARPES measurements. Our experimental discovery of hidden altermagnetism broadens the scope of unconventional magnetism and opens routes to exploring emergent phenomena from real-space modulations of altermagnetic order.
\end{abstract}

\maketitle

\par Altermagnetism is characterized by a collinear antiferromagnetic (AFM) structure with zero net magnetization, yet exhibits momentum-dependent spin polarization in the electronic band structure \cite{DefAMPRX,ProposeAMPRX}. In altermagnets, the spin-up and spin-down sublattices cannot be connected by either translation or inversion operations, and thus the “spin-flip-translation” ($\tau$T) and “time-reversal-space-inversion” (PT) symmetries are both broken, leading to spontaneous spin splittings in momentum space. Following extensive theoretical studies \cite{DefAMPRX,ProposeAMPRX,PhysRevB.75.115103,AMpropose2019,AMpropose2020SA,PRB101AMpropose,AMpropose2020PRB,AMproposeFeSb22021,CpairedSVL2021NC,GMRTMRcalPRX,spingroupPRX,spincurrentPRL,AM2DstackPRL,vdWAMPRL,PhysRevB.109.024404,AMoctupoles,AFMreview}, altermagnetism has been experimentally reported in several materials, including RuO$_2$ \cite{spincurrentexpRuONE,AHEexpRuONE,SSTexpRuOPRLA,SSTexpRuOPRLB,SCCASSEexpPRL,RuO2APLAHE,RuO2ARPESSA}, MnTe \cite{AHEexpMnTePRL,MnTeARPESNature,MnTeARPESPRL,MnTeXMCD,MnTeARPESPRB,RN170}, CrSb \cite{CrSbfilmARPESNC,RN93,RN89,RN95,CrSbtopoARPES,CrSbnanolet,RN204,RN205}, Mn$_5$Si$_3$ \cite{RN129,Mn5Si3AHEPRB}, KV$_{2}$Se$_{2}$O \cite{Jiang2025}, and Rb$_{1-\delta}$V$_{2}$Te$_{2}$O \cite{Zhang2025}. While two-dimensional $d$-wave altermagnets are favorable for generating spin currents \cite{CoulombDrag}, their identifications in real materials remain somewhat controversial: in RuO$_2$, the existence of altermagnetism is still under debate \cite{PhysRevLett.132.166702,RN128}; the G-type AFM recently discovered in KV$_{2}$Se$_{2}$O seems inconsistent with its previously reported $d$-wave altermagnetism \cite{27n8-5q4l}. These debates underscore the necessity of combining neutron scattering with angle-resolved photoemission spectroscopy (ARPES) to probe both the magnetic structures and spin-resolved band structures of candidate altermagnets.

\begin{figure}[tp]
	\includegraphics[width=1.0\columnwidth]{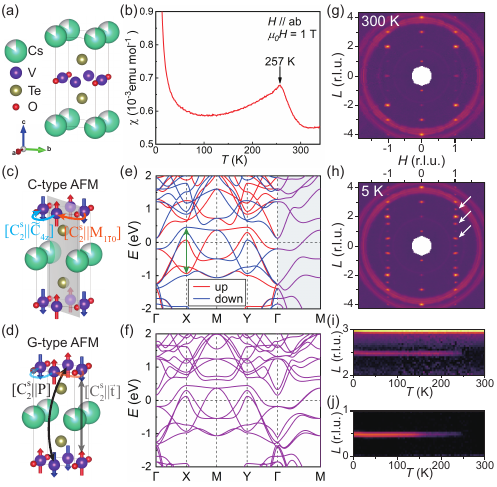}
	\centering
	\caption{Magnetic ground state and calculated electronic structure of Cs$_{1-\delta}$V$_2$Te$_2$O.
		(a) Crystal structure.
		(b) Temperature-dependent magnetic susceptibility.
		(c,d) Two possible AFM states with lowest energies: C-type (c) and G-type AFM (d). The corresponding symmetry operations are indicated.
		(e,f) Calculated spin-resolved band structures for C-type (e) and G-type AFM (f).
		(g,h) Neutron diffraction intensity maps in the $[H0L]$-plane at $T$ = 300~K (g) and 5~K (h). Magnetic peaks with half-integer $L$ are indicated in (h) using white arrows.
        (i,j) Temperature dependence of the $\mathbf{Q}=(0,-1,\frac{5}{2})$ (i) and $(0,-1,\frac{1}{2})$ (j) magnetic peaks. }
		\label{fig1}
\end{figure}

\par In principle, a two-dimensional altermagnets can vertically stack in a fashion that preserves PT symmetry, yielding spin degeneracy in momentum space yet layer-dependent altermagnetic spin patterns. This is inspired by the previous studies showing that relativistic spin-orbit coupling (SOC) in  certain centrosymmetric crystals can produce substantial local spin polarization, termed ``hidden spin polarization'' \cite{RN203,nanolet2013}. In such systems, the bulk electronic bands remain spin-degenerate in momentum space due to global inversion symmetry, but the electronic states can carry sizable spin polarizations localized on specific atomic sites or layers that locally break inversion symmetry. Such hidden spin polarization can be detected through surface-sensitive spin-resolved ARPES measurements \cite{NP2014exp,RN213,RN207,Yao2017,PhysRevB101035102,RN208,Gatti2021,PhysRevB.104.035145}, and its significance is further amplified in PT-symmetric magnets, for enabling the electrical control of the N\'{e}el order \cite{PhysRevLett.113.157201,doi:10.1126/science.aab1031,PhysRevLett.129.276601}. By extending the idea of hidden spin polarization to altermagnetism (which does not require SOC), ``hidden altermagnets'' with locally spin-polarized states have been theoretically proposed \cite{hiddenpolarnoSOC,hiddenAMPRL,hiddenAMFP}, although they have not been observed experimentally so far. 

\par In this work, we report the first observation of hidden altermagnetism, in Cs$_{1-\delta}$V$_2$Te$_2$O single crystals \cite{CVTOarXiv}. We combine bulk-sensitive neutron scattering, surface-sensitive ARPES and spin-resolved ARPES, and density functional theory (DFT) calculations to obtain a consistent understanding of both magnetic structures and spin-polarized electronic states. Our Neutron diffraction establishes a G-type AFM structure that preserves the spin degeneracy of bulk bands in momentum space, while ARPES measurements reveal two-dimensional bands with momentum-dependent local spin polarizations. Our experimental discovery of hidden altermagnetism broadens the scope of altermagnetism and open up the opportunities to explore emergent phenomena associated with real-space modulations of local altermagnetic order. 

\begin{figure*}[tp]
	\includegraphics[width=1.8\columnwidth]{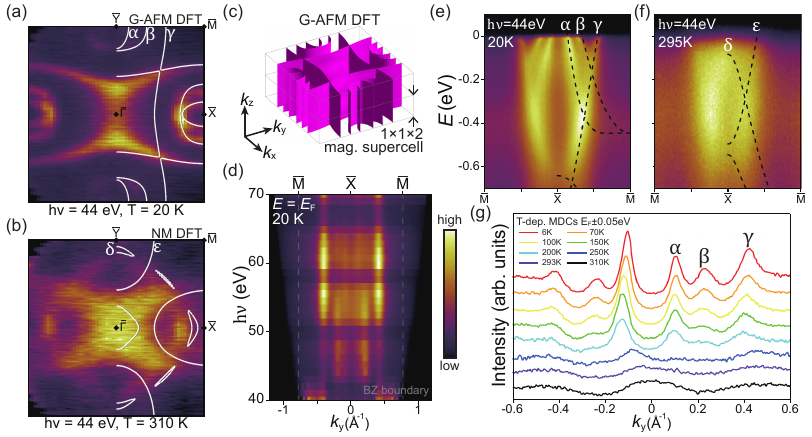}
	\centering
	\caption{Electronic structure and temperature evolution of Cs$_{1-\delta}$V$_2$Te$_2$O from ARPES.
		(a,b) The $k_x$-$k_y$ FS taken with 44~eV photons at $T$ = 20~K (a) and 310~K (b), respectively. The calculated Fermi contours at $k_z$ = 0 (white curves) are overlaid on the right half. 
		(c) A three-dimensional view of the calculated FS corresponding to the G-type AFM.
		(d) The photon-energy-dependent scan at $E_F$ along $\bar{M}-\bar{X}-\bar{M}$.
		(e,f) Energy-momentum cuts along $\bar{M}-\bar{X}-\bar{M}$ at $T$ = 20~K (e) and 295~K (f), respectively. The calculated band structures for the G-type AFM and nonmagnetic states (black dashed curves) are overlaid on the right half.
        (g) Temperature-dependent MDCs at $E_F$ along $\bar{M}-\bar{X}-\bar{M}$, taken with 21.2~eV photons. Original data can be found in Fig. S5 \cite{supplementary}. 
        \label{fig2}
	}
\end{figure*}

\par Cs$_{1-\delta}$V$_2$Te$_2$O ($\delta$ $\approx$ 0.23) is a quasi-two-dimensional compound with alternating Cs and Te$_2$V$_2$O building blocks, crystallizing in the $P4/mmm$ (No. 123) space group [Fig. \ref{fig1}(a)] \cite{CVTOarXiv}. Details of sample growth and measurements can be found in \cite{supplementary}. Temperature-dependent magnetic susceptibility $\chi$ shows a clear AFM transition with a N\'{e}el temperature $T_N$ $\approx$ 257~K [Fig. \ref{fig1}(b)]. DFT calculations of different magnetic configurations further show that a N\'{e}el-type spin alignment within the ab-plane has the lowest energy (Fig. S1 in \cite{supplementary}), although it is difficult to distinguish whether the C-type or G-type AFM [Figs. \ref{fig1}(c) and \ref{fig1}(d)] is the ground state from DFT, due to their nearly degenerate energies. 

\par The magnetic structure has a profound impact over the spin-space group symmetry and the corresponding spin splittings in momentum space. For C-type AFM [ferromagnetic alignment between adjacent layers, Fig. \ref{fig1}(c)], the spin-opposite sublattices can be linked only by rotation $[C_2^{s}\|C_{4z}]$ or mirror $[C_2^{s}\|M_{1\bar{1}0}]$ operations, leading to broken $\tau$T and PT symmetries and hence planar-type $d$-wave altermagnetism \cite{DefAMPRX,CpairedSVL2021NC}. Here $[R_i\|R_j]$ denotes spin-space group operations, where $R_i$ ($R_j$) acts only on spin (real) space, respectively \cite{SG1966,SG1974,PhysRevX.14.031037,PhysRevX.14.031038,PhysRevX.14.031039}. Such an altermagnetic state was recently reported in KV$_2$Se$_2$O \cite{Jiang2025} and Rb$_{1-\delta}$V$_2$Te$_2$O \cite{Zhang2025}. In contrast, for G-type AFM [AFM alignment between adjacent layers, Fig. \ref{fig1}(d)], the spin-opposite sublattices can be connected by translation $[C_2^{s}\|\vec{t_z}]$ or inversion $[C_2^{s}\|P]$ operations, resulting in spin-degenerate bulk bands. Such spin splittings or degeneracies are verified by DFT calculations [Figs. \ref{fig1}(e) and \ref{fig1}(f)]: while the C-type AFM state exhibits large spin splittings along $\Gamma-X(Y)-M$, expected for a $d$-wave altermagnet, the G-type AFM state shows no bulk spin polarization. Interestingly, if interlayer hoppings can be ignored, the G-type AFM can be viewed as stacked $d$-wave altermagnetic layers with alternating local spin polarizations - this is essentially the hidden altermagnet \cite{hiddenAMPRL,hiddenAMFP}, to be discussed below. 

\par To probe the magnetic structure, we performed single-crystal neutron diffraction measurements. Comparing $[H0L]$ diffraction maps above and below $T_N$ [Figs. \ref{fig1}(g) and \ref{fig1}(h)] reveals additional reflections at $(1,0,1/2)$ and symmetry-related positions that appear only in the AFM phase, as further confirmed by the detailed temperature-dependent measurements [Figs. \ref{fig1}(i) and \ref{fig1}(j)]. The half-integer-$L$ peaks are characteristic of a G-type AFM structure, thereby ruling out the C-type AFM order. The G-type AFM with spins oriented along the $c$-axis is further established by fitting the neutron diffraction data at 5~K using \texttt{JANA2020} (Fig. S2 in \cite{supplementary}), revealing an ordered moment of $M=1.03(7)~\mu_{\rm B}$ per V atom.

\begin{figure*}[tp]
	\includegraphics[width=2.0\columnwidth]{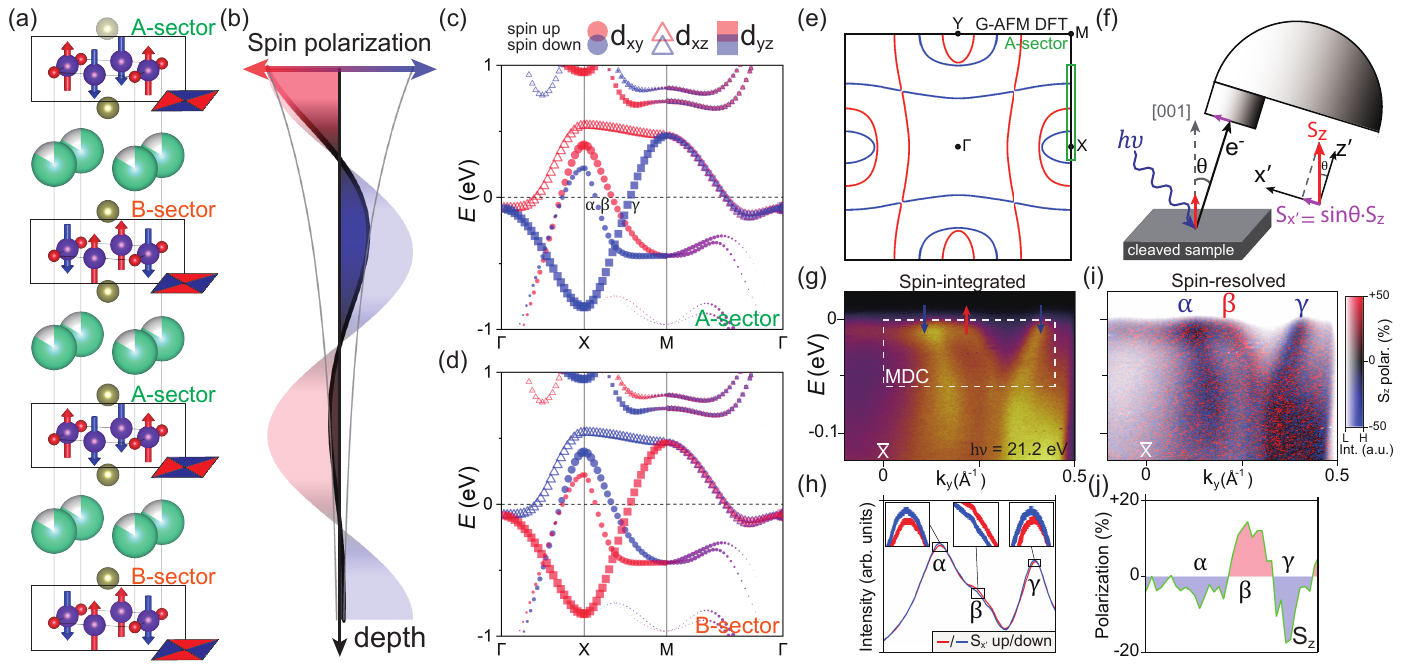}
	\centering
	\caption{Detecting the local spin polarization by spin-resolved ARPES.
		(a) A schematic of hidden altermagnetism in Cs$_{1-\delta}$V$_2$Te$_2$O.  
        (b) Local spin signals, detected by spin-resolved ARPES, decay exponentially with $z$ and oscillate due to hidden altermagnetism. This plot shows the case of A-sector terminated surface.
		(c,d) Projected spin polarization and orbital weight of the calculated bands at the A (c) and B (d) sectors. 
		(e) Calculated two-dimensional FS for the A sector.  
        (f) Geometry of the spin-resolved ARPES measurement. The spin polarization along $z$, $S_z$, is equal to $S_{x'}$/sin($\theta$), where $S_{x'}$ is the effective spin polarization along $x'$ measured by spin-resolved ARPES.
        (g) Spin-integrated spectrum (up + down) along $\bar{X}-\bar{M}$ as indicated in (e). 
        (h) The spin-resolved MDCs integrated over the white dashed rectangle defined in (g). Insets are zoom-in views of each band. 
        (i) Spin-resolved spectrum (up $-$ down) corresponding to (g). The red or blue color indicates positive or negative $S_z$ spin polarization, while transparent or opaque indicates low or high intensity, as indicated by the color-opacity bar on the right. 
        (j) Momentum-dependent spin polarization $S_z$ derived from (h).
        The data in (g-j) are obtained from the same sample with mostly A-sector terminated surface. 
        \label{fig3}
	}
\end{figure*}

\par To probe the electronic states, we performed ARPES measurements on single crystals cleaved $in$-$situ$, whose surfaces consist of both Cs-layer and Te-layer terminated regions (Fig. S3 in \cite{supplementary}). Both terminations yield similar Fermi surface (FS) experimentally and hence the in-plane N\'{e}el-type AFM order is most likely preserved at the surface, consistent with DFT slab calculations. Figures \ref{fig2}(a) and \ref{fig2}(b) show the experimental FS in the $k_x$-$k_y$ plane below and above $T_N$. In the G-type AFM state [Fig. \ref{fig2}(a)], the FS consists of two sets of hole pockets ($\alpha$ and $\beta$) centered at $\bar{X}$/$\bar{Y}$ points, and quasi-one-dimensional electron bands ($\gamma$) along the $k_x$/$k_y$ directions. The experimental FS can be well captured by the DFT calculations shown in Fig. \ref{fig2}(a). Figure \ref{fig2}(c) shows a three-dimensional view of the calculated FS in the G-type AFM state, which shows very little $k_z$ dispersion due to weak interlayer couplings. Such a two-dimensional FS is directly verified by photon-energy-dependent ARPES measurements in Fig. \ref{fig2}(d) (additional data shown in Fig. S4 \cite{supplementary}), where straight Fermi contours with negligible $k_z$ dispersion can be identified. 

\par Upon warming above $T_N$, the experimental FS undergoes dramatic changes [Fig. \ref{fig2}(b)]: the FS now consists of a large diamond-shaped pocket ($\delta$) centered at $\bar{\Gamma}$ and elliptical electron pockets ($\epsilon$) centered at $\bar{X}$/$\bar{Y}$ points. Such an electronic reconstruction can be further revealed by the energy-momentum cuts shown in Figs. \ref{fig2}(e) and \ref{fig2}(f): the $\alpha$ and $\beta$ bands at 20~K merge to form the $\delta$ band with lower energy above $T_N$, while the $\gamma$ band evolves into the $\epsilon$ band. The experimental band structure above $T_N$ is in good agreement with the nonmagnetic DFT calculations shown in Figs. \ref{fig2}(b) and \ref{fig2}(f), although some fine features from DFT calculations cannot be resolved experimentally due to thermal broadening. Figure \ref{fig2}(g) shows a detailed temperature evolution of the momentum distribution curves (MDCs) at $E_F$ (additional data and analysis shown in Fig. S5 \cite{supplementary}), which reveals that the splitting between the $\alpha$ and $\beta$ bands gradually decreases with increasing temperature, eventually becoming indistinguishable near 250 K, close to $T_N$. We also performed careful temperature cycles for the ARPES measurements, which rule out extrinsic effects due to sample aging (Fig. S6 in \cite{supplementary}). 

\par The G-type AFM order in Cs$_{1-\delta}$V$_2$Te$_2$O can be viewed as a vertical stack of two-dimensional altermagnetic layers with alternating local spin polarizations [Fig. \ref{fig3}(a)], which we label as A- and B-sectors, respectively. Although the global inversion symmetry preserves the spin degeneracy of bands in momentum space (assuming very delocalized electrons), the negligible $k_z$ dispersion allows for development of hidden altermagnetism: the electrons are effectively localized along $z$ and move only within each layer, leading to local and opposite spin polarizations in A- and B-sectors, respectively. To see this, we project the calculated bands into the V $3d$ orbitals located at A- or B-sector [Figs. \ref{fig3}(c) and \ref{fig3}(d)], which make dominant contributions to the FS [Fig. \ref{fig3}(e) shows the A-sector FS]. The projections reveal local altermagnetic splittings with opposite spin polarizations for A- and B-sectors. The calculations further reveal a local spin polarization close to 100\% near $E_F$ at the center of each sector, due to the very weak interlayer hopping. Interestingly, the spin splitting for the $\gamma$ band is accompanied by a large orbital polarization \cite{CVTOarXiv}: along $\Gamma-X-M$, the spin-up and spin-down $\gamma$ bands at the A-sector are dominated by {$d_{xz}$ and $d_{yz}$} orbitals, respectively [Fig. \ref{fig3}(c)]. {As for the B-sector, the spin polarization reverses while the orbital character remains unchanged} [Fig. \ref{fig3}(d)]. Such a pronounced ``spin-orbital locking'', caused by spin-dependent Coulomb interaction (instead of relativistic SOC), is confirmed by the polarization-dependent ARPES measurements (Fig. S8 in \cite{supplementary}). 

\par Although the opposite local spin polarizations of A- and B- sectors lead to compensated spin currents, surface-sensitive measurements, e.g., spin-resolved ARPES, are capable of probing the local spin polarization - such an approach has been successfully used to detect the SOC-driven hidden spin polarization \cite{NP2014exp,RN213,RN207,Yao2017,PhysRevB101035102,RN208,Gatti2021,PhysRevB.104.035145}. Figure \ref{fig3}(b) illustrates how surface-sensitive spin-resolved ARPES can detect the hidden altermagnetism in Cs$_{1-\delta}$V$_2$Te$_2$O: the exponential decay of photoelectron signals along $z$, caused by a small mean free path $\sim$ 1~nm, guarantees dominant contributions from the top sector (the interlayer distance is 0.89~nm). A simple estimate yields a maximal spin polarization of $\sim$71\% in spin-resolved ARPES for a purely A- or B-sector terminated surface (section V in \cite{supplementary}). To probe the local spin polarization, we performed spin-resolved ARPES measurements along $\bar{X}-\bar{M}$ as indicated in Fig. \ref{fig3}(e) [measurement geometry is shown in Fig. \ref{fig3}(f)]. While the spin-integrated ARPES spectrum [Fig. \ref{fig3}(g)] reveals the expected $\alpha$, $\beta$ and $\gamma$ bands, the corresponding spin-resolved MDCs near $E_F$ [Fig. \ref{fig3}(h)] confirm the negative-positive-negative spin polarization predicted from DFT (additional data shown in Fig. S9 in \cite{supplementary}). Based on the calibrated Sherman function from Au(111), we obtain the spin-resolved ARPES spectrum in Fig. \ref{fig3}(i), which further verifies the expected spin polarizations of $\alpha$, $\beta$ and $\gamma$ bands, respectively. The extracted spin polarization $S_z$, shown in Fig. \ref{fig3}(j), is up to $\sim$15\% for $\beta$ and $\gamma$ bands.

\par Note that the maximal spin polarization obtained experimentally is much smaller than the theoretical {maximum} ($\sim$71\%). This reduction arises from the large beam spot in spin-resolved ARPES, leading to a partial cancellation in signals from A and B sectors (Fig. S11 in \cite{supplementary}). Under identical measurement geometry, we {can observe} both negative-positive-negative and positive-negative-{positive} spin patterns along $\bar{X}-\bar{M}$, which correspond to samples with mainly A-sector (Fig. \ref{fig3}) and B-sector (Fig. S10 in \cite{supplementary}) terminated surfaces, respectively. Furthermore, a reversal of spin polarization is observed for the same sample upon a 90$^{\circ}$ in-plane rotation (Fig. S10 in \cite{supplementary}), confirming the in-plane $d$-wave symmetry of spin splitting. 

\par Compared with traditional altermagnetism, our discovery of hidden altermagnetism opens up the opportunities to explore emergent phenomena from the real-space modulation of altermagnetism. For example, electric gating along $\pm z$ can convert a layered hidden altermagnet into a global altermagnet with tunable and reversible spin polarizations, as shown in our DFT calculations in Fig. S12 of \cite{supplementary}. Significantly, such a property opens another possibility for electric switchable altermagnetism \cite{PhysRevLett.134.106802}, favorable for the application of electric-field controllable spin-filtering tunnel junction without the need of switching the N\'{e}el vector. In addition, for superconductors driven by altermagnetic spin fluctuations \cite{Mazin2025,Lu2025,Ma2025}, the opposite sign of local spin splittings in hidden altermagnet could lead to odd-parity pairing states, as observed in locally non-centrosymmetric heavy fermion superconductor CeRh$_2$As$_2$ \cite{Khim2021science}. 

\par In summary, by combining neutron scattering, ARPES measurements and DFT calculations, we report the experimental discovery of hidden altermagnetism in Cs$_{1-\delta}$V$_2$Te$_2$O, where the altermagnetic-like band splittings with local spin polarizations can be observed. Our discovery bridges the gap between conventional PT-symmetric AFM and altermagnetism, and thus broadens the scope of unconventional magnetism \cite{NPunconventionalAFM}. Furthermore, our work highlights the importance of combining measurements of both magnetic and electronic structures, in order to identify and distinguish the global/local spin splittings in different types of unconventional magnets. 

\par Note: During the review process of this paper, we became aware of a neutron scattering study on Rb$_{1-\delta}$V$_{2}$Te$_{2}$O \cite{xie2026GAFMRVTO}, which also reports the G-type AFM order. Taken together, all three V-based 1221 compounds studied {to date} (KV$_{2}$Se$_{2}$O, Cs$_{1-\delta}$V$_2$Te$_2$O and Rb$_{1-\delta}$V$_{2}$Te$_{2}$O) show the G-type AFM order with appreciable spin polarizations in the electronic states, consistent with the hidden altermagnetism proposed in this work. Hidden altermagnetism can be further verified by spin-polarized scanning tunneling microscopy (STM). Very recent STM studies have indeed revealed $d$-wave-like spin-split states on flat surfaces \cite{Wangzhenyu2025STMKVSO,WeiLi2025STMCVSO,JL2026STMKVSO}. It would be interesting in the future to confirm the opposite spin polarizations between adjacent sectors (along $z$) separated by a single step, as expected for a hidden altermagnet.

\par This work is supported by the National Key R$\&$D Program of China (Grant No.~2023YFA1406303, No.~2022YFA1402200), the National Natural Science Foundation of China (No.~12525408, No.~12350710785, No.~12174331, No.~12274363, No.~62427901, No.~11927807) and the Strategic Priority Research Program of the Chinese Academy of Sciences (No.~XDB0670000). We thank the Shanghai Synchrotron Radiation Facility (SSRF) of BL03U (31124.02.SSRF.BL03U) for the assistance on ARPES measurements. A portion of this research used resources at the Spallation Neutron Source, a DOE Office of Science User Facility operated by the Oak Ridge National Laboratory. The beam time was allocated to TOPAZ on Proposal No.~IPTS-34478. X. W. acknowledges research sponsored by the Laboratory Directed Research and Development Program of Oak Ridge National Laboratory, managed by UT Battelle, LLC, for the U.S. Department of Energy. We would like to thank {Qi Hu,} Zhanfeng Liu, Jiayu Liu, Yu Huang, Yiwei Cheng, Tongrui Li, Hui Tian, Zongyi Wang and Yunbo Wu for experimental help. We are also grateful to Prof. Gabriel Aeppli, Prof. Yuanfeng Xu, Prof. Lin Jiao and Prof. Huiqiu Yuan for helpful discussions. Some of the images in the paper were created using VESTA software \cite{vesta}.

\nocite{CVTOarXiv,Coates2018,NeuXtalViz,Reshniak2024,Schultz2014,Sheldrick2015,Petek2023,RN131,10.1063/5.0142548,VASP1993,PAW1999,PBE1996}


%

\end{document}